\documentstyle[12pt]{article}

\newcounter{eq}
\newcounter{sc}

\def\overleftrightarrow#1{\vbox{\ialign{##\crcr
 $\leftrightarrow$\crcr\noalign{\kern-1pt\nointerlineskip}
 $\hfil\displaystyle{#1}\hfil$\crcr}}}

\newlength{\minitwocolumn}
\begin{document}

\begin{flushright}
DPUR/TH/66\\
March , 2020\\
\end{flushright}
\vspace{20pt}

\pagestyle{empty}
\baselineskip15pt

\begin{center}
{\large\bf  Spontaneous Symmetry Breakdown of Scale Symmetries
\vskip 1mm }

\vspace{20mm}

Ichiro Oda\footnote{
           E-mail address:\ ioda@sci.u-ryukyu.ac.jp
                  }

\vspace{10mm}
           Department of Physics, Faculty of Science, University of the 
           Ryukyus,\\
           Nishihara, Okinawa 903-0213, Japan\\

\end{center}


\vspace{10mm}
\begin{abstract}

We discuss spontaneous symmetry breakdown (SSB) of both global and local scale symmetries in
scalar-tensor gravity with two scalar fields, one of which couples nonminimally to scalar curvature
while the other is a normal scalar field. 
In case of a global scale symmetry, by moving from the Jordan frame to the Einstein frame,
a normal scalar field becomes massive while the dilaton remains massless after the SSB. 
In case of a local scale symmetry, we take a gauge fixing condition for the local scale invariance, 
$a R + b \phi^2 = k$, which was found in our previous study of a Weyl's quadratic gravity. Together
with locally scale invariant potential terms in a classical action, this gauge condition generates 
a Higgs potential whose vacuum expectation value (VEV) produces the Einstein-Hilbert action 
with an $R^2$ term in the lowest level of approximation. 
One interesting aspect in this model is that a massless Nambu-Goldstone boson associated with 
a scale invariance is absorbed into the metric tensor field and consequently an $R^2$ term 
is induced in the action.

\end{abstract}

\newpage
\pagestyle{plain}
\pagenumbering{arabic}


\section{Introduction}

In this article, we wish to discuss how both global and local scale symmetries are spontaneously broken
in a scale invariant scalar-tensor gravity with two scalar fields, one of which is a dilaton and 
couples nonminimally to scalar curvature \cite{Fujii}. The theory has thus far played an important role 
when we discuss a manifestly scale invariant regularization method \cite{Englert}.

There are several motivations behind the present study. One motivation comes from a very
simple question: Starting with a conformally invariant scalar-tensor gravity,\footnote{Following 
the textbook by Fujii and Maeda  \cite{Fujii}, we shall call a global scale symmetry simply ``scale 
symmetry'' whereas we call a local scale symmetry ``conformal symmetry'' in this article.} 
is it possible to construct a physically viable and interesting theory without taking a ``unitary gauge'' 
where the ``dilaton'' is put to be a constant, $\phi (x) = const.$? With the unitary gauge for conformal
symmetry, there is no ghost associated with the scalar field $\phi (x)$, but then we cannot
have any benefit of conformal symmetry, i.e., only the Einstein-Hilbert action with a positive
Newton constant, which is not invariant under conformal transformation, is left behind. 
Thus, conformal symmetry in the conformally invariant scalar-tensor gravity is sometimes
called $\textit{fake symmetry}$ \cite{Jackiw, Oda1}, but is it really useless?  In this study, 
we show that it is not the case, but the conformally invariant scalar-tensor gravity produces
the Einstein-Hilbert action plus an $R^2$ term in the lowest level of approximation in fields,
and a resultant action is free from a ghost associated with the dilaton via the SSB when
we take an appropriate gauge fixing condition for conformal symmetry.  

As a second motivation, in our previous work \cite{Oda2}, starting with a quadratic gravity 
in the Weyl conformal geometry \cite{Weyl, O'Raifeartaigh}, we have obtained the Einstein-Hilbert action 
with a positive cosmological constant and a massive Weyl gauge field by means of a gauge
condition for the Weyl gauge symmetry, $a \tilde R + b \phi^2 = k$, where $\tilde R$ is 
Weyl's scalar curvature and $a, b, k$ are constants. A question then arises if this argument
could be inverted or not, i.e., beginning with the conformally invariant scalar-tensor gravity
with a linear scalar curvature, is it possible to get a quadratic gravity (plus the Einstein-Hilbert
action) through an analogous gauge condition, $a R + b \phi^2 = k$, for conformal symmetry.
We will also see that this is indeed the case.

The third motivation is that even if we have presented a new idea of the SSB of the Weyl gauge
symmetry based on the Weyl geometry \cite{Oda2}, its content is a bit difficult to understand 
due to the complicated structure of the Weyl geometry. Thus, we wish to account for our idea
more clearly in the framework of a simpler theory, which is nothing but the conformally invariant 
scalar-tensor gravity in the Riemann geometry.

The final motivation is related to our conjecture of the origin of the Higgs potential:
The Higgs potential might stem from the gauge fixing condition for conformal symmetry or
the Weyl gauge symmetry \cite{Oda2}. As regards this conjecture, let us recall that
in the standard model (SM) the Higgs potential for symmetry breaking of gauge symmetries is 
introduced by hand without asking its origin, and the SM has a classical scale symmetry if one sets 
a (negative) Higgs mass term to zero \cite{Bardeen}. This situation might support our conjecture 
that the Higgs potential could be generated by symmetry breaking of conformal symmetry.

We close this section with an overview of this article. In Section 2, we briefly review the SSB
of scale symmetry in a scale invariant scalar-tensor gravity with two scalar fields.
In Section 3, we work with the conformally invariant scalar-tensor gravity. We see that
with a suitable gauge fixing condition for conformal symmetry, together with conformally
invariant potential terms in a classical action, the Higgs potential is generated, and as a result 
the Einstein-Hilbert action plus an $R^2$ term is generated by the Higgs mechanism. 
Section 4 is devoted to the conclusion.

\section{Spontaneous symmetry breakdown of global scale symmetry} 

There is a well-known mechanism of spontaneous symmetry breakdown of a global scale symmetry
\cite{Fujii, Oda3}. In this section, as a prelude to the conformally invariant scalar-tensor gravity
treated in the next section, we shall briefly review a scale invariant scalar-tensor gravity with two scalar 
fields, explain how the scale symmetry is broken spontaneously, and then point out unsatisfactory 
points of this SSB mechanism.

As a model of a scale invariant scalar-tensor gravity with two scalar fields, let us work with the following
Lagrangian density in the Jordan frame\footnote{We follow the conventions and notation adopted
in the MTW textbook \cite{MTW}.}:
\begin{eqnarray}
{\cal{L}} = \sqrt{-g} \left( \frac{1}{2} \xi \phi^2 R - \frac{1}{2} \epsilon g^{\mu\nu} \partial_\mu \phi
\partial_\nu \phi -  \frac{1}{2} g^{\mu\nu} \partial_\mu \Phi \partial_\nu \Phi 
- \frac{\lambda_1}{4} \phi^4 - \frac{\lambda_2}{2} \phi^2 \Phi^2 - \frac{\lambda_3}{4} \Phi^4 \right),
\label{J-model}
\end{eqnarray}
where $\xi$ is a constant, and $\epsilon$ takes the value $+1$ for $\phi$ being a normal field while it does $-1$
for $\phi$ being a ghost field. Moreover, $\phi$ and $\Phi$ are two distinct scalar fields, and $\lambda_i  (i = 1, 2, 3)$
are dimensionless coupling constants.  As often taken in the application for the BSM \cite{Ghilencea}, we assume that 
$\lambda_1 > 0, \lambda_3 > 0$ and $\lambda_2 < 0$, and furthermore $| \lambda_2 | \ll \lambda_1, \lambda_3 \approx
{\cal {O}} (0.1)$. The conformally invariant scalar-tensor gravity, which will be considered in the next section, 
corresponds to either the case of $\xi = \frac{1}{6}$ and $\epsilon = - 1$ or the case of $\xi = - \frac{1}{6}$ 
and $\epsilon = 1$.  In this section, since we consider only a globally scale invariant theory, we assume 
$\xi > 0$ and $6 + \frac{\epsilon}{\xi} > 0$.

From this Lagrangian density, it is straightforward to derive the field equations for the metric tensor $g_{\mu\nu}$ 
and the two scalar fields $\phi, \Phi$ whose result is written as
\begin{eqnarray}
&{}& 2 \varphi G_{\mu\nu} + 2 ( g_{\mu\nu} \Box - \nabla_\mu \nabla_\nu ) \varphi = T_{\mu\nu},
\nonumber\\
&{}& \xi \phi R + \epsilon \Box \phi - \lambda_1 \phi^3 - \lambda_2 \phi \Phi^2 = 0,
\nonumber\\
&{}& \Box \Phi - \lambda_2 \phi^2 \Phi - \lambda_3 \Phi^3 = 0,
\label{Field-eq}
\end{eqnarray}
where we have defined 
\begin{eqnarray}
\varphi &=& \frac{1}{2} \xi \phi^2, \qquad G_{\mu\nu} = R_{\mu\nu} - \frac{1}{2} g_{\mu\nu} R,
\qquad
\Box \varphi = \frac{1}{\sqrt{-g}} \partial_\mu ( \sqrt{-g} g^{\mu\nu} \partial_\nu \varphi ),
\nonumber\\
T_{\mu\nu} &=& \epsilon \partial_\mu \phi \partial_\nu \phi + \partial_\mu \Phi \partial_\nu \Phi
+ g_{\mu\nu} \Biggl( - \frac{1}{2} \epsilon g^{\alpha \beta} \partial_\alpha \phi \partial_\beta \phi  
-  \frac{1}{2} g^{\alpha \beta} \partial_\alpha \Phi \partial_\beta \Phi 
\nonumber\\
&-&\frac{\lambda_1}{4} \phi^4 - \frac{\lambda_2}{2} \phi^2 \Phi^2 - \frac{\lambda_3}{4} \Phi^4 \Biggr).
\label{Various def}
\end{eqnarray}
Using these field equations, one can derive the following equation:
\begin{eqnarray}
\Box ( \varphi + \frac{\zeta^2}{2} \Phi^2 ) = 0,
\label{Derived-eq}
\end{eqnarray}
where we have defined $\zeta^{-2} \equiv 6 + \frac{\epsilon}{\xi} > 0$.

The key step for the SSB of scale invariance is to move from the Jordan frame (J-frame) to 
the Einstein frame (E-frame) by applying a conformal transformation, i.e., a local scale transformation:
\begin{eqnarray}
g_{\mu\nu} \rightarrow g^\prime_{\mu\nu}  = \Omega^2 (x) g_{\mu\nu}, \qquad
\phi \rightarrow \phi^\prime = \Omega^{-1} (x) \phi, \qquad
\Phi \rightarrow \Phi^\prime = \Omega^{-1} (x) \Phi.
\label{Conformal transf}
\end{eqnarray}
After some calculations, we can derive the transformation rule for scalar curvature \cite{Fujii}:
\begin{eqnarray}
R = \Omega^2 (x) ( R^\prime + 6 \Box^\prime f - 6 g^{\prime \mu\nu} f_\mu f_\nu ),
\label{R-transf}
\end{eqnarray}
where we have defined
\begin{eqnarray}
f = \log \Omega, \qquad f_\mu = \partial_\mu f, \qquad
\Box^\prime f = \frac{1}{\sqrt{-g^\prime}} \partial_\mu ( \sqrt{-g^\prime} g^{\prime \mu\nu}
\partial_\nu f ).
\label{Def-f}
\end{eqnarray}
Using these relations, we find that the Lagrangian density (\ref{J-model}) can be cast
to the form in a new conformal frame:
\begin{eqnarray}
{\cal{L}} &=& \sqrt{-g^\prime} \Biggl[ \frac{1}{2} \xi \phi^{\prime 2} ( R^\prime + 6 \Box^\prime f 
- 6 g^{\prime \mu\nu} f_\mu f_\nu )
- \frac{1}{2} \epsilon \Omega^{-2} g^{\prime \mu\nu} \partial_\mu ( \Omega \phi^\prime )
\partial_\nu  ( \Omega \phi^\prime )
\nonumber\\
&-&  \frac{1}{2} \Omega^{-2} g^{\prime \mu\nu} \partial_\mu ( \Omega \Phi^\prime )
\partial_\nu  ( \Omega \Phi^\prime )
- \frac{\lambda_1}{4} \phi^{\prime 4} - \frac{\lambda_2}{2} \phi^{\prime 2} \Phi^{\prime 2}
- \frac{\lambda_3}{4} \Phi^{\prime 4} \Bigr].
\label{New-frame}
\end{eqnarray}

Moving to the E-frame requires us to choose the scalar field $\phi^\prime$ to\footnote{In case of
conformal symmetry, this condition is called the ``Einstein gauge'' or ``unitary gauge''.}
\begin{eqnarray}
\phi^\prime = \frac{M_{Pl}}{\sqrt{\xi}},
\label{E-frame1}
\end{eqnarray}
where  $M_{Pl}$ is the (reduced) Planck mass defined as $M_{Pl} = \frac{1}{\sqrt{8 \pi G}} = 2.44 
\times 10^{18} GeV$ with $G$ being the Newton constant.
Then, in the E-frame, up to a total derivative, the Lagrangian density (\ref{New-frame}) reduces to the form:
\begin{eqnarray}
{\cal{L}} = \sqrt{-g^\prime} \left( \frac{M_{Pl}^2}{2} R^\prime - \frac{1}{2} g^{\prime \mu\nu} 
\partial_\mu \sigma \partial_\nu \sigma - \frac{1}{2} g^{\prime \mu\nu} {\cal{D}}_\mu \Phi^\prime 
{\cal{D}}_\nu \Phi^\prime - \frac{\lambda_1}{4} \frac{M_{Pl}^4}{\xi^2}  
- \frac{\lambda_2}{2} \frac{M_{Pl}^2}{\xi} \Phi^{\prime 2} - \frac{\lambda_3}{4} \Phi^{\prime 4} \right).
\label{E-model}
\end{eqnarray}
Here we have defined
\begin{eqnarray}
\Omega (x) = e^{\frac{\zeta}{M_{Pl}}  \sigma (x)}, \qquad
{\cal{D}}_\mu \Phi^\prime = \left( \partial_\mu + \frac{\zeta}{M_{Pl}} 
\partial_\mu \sigma \right) \Phi^\prime,
\label{Dilaton}
\end{eqnarray}
where a scalar field $\sigma$ is called "dilaton". 

Now, owing to our assumption $\lambda_1 > 0, \lambda_3 > 0$ and $\lambda_2 < 0$, we have a Higgs potential 
given by
\begin{eqnarray}
V(\Phi^\prime) &=& \frac{\lambda_3}{4} \Phi^{\prime 4}   
+ \frac{\lambda_2}{2} \frac{M_{Pl}^2}{\xi} \Phi^{\prime 2} + \frac{\lambda_1}{4} \frac{M_{Pl}^4}{\xi^2}
\nonumber\\
&=& \frac{\lambda_3}{4} \left( \Phi^{\prime 2} - \frac{|\lambda_2|}{\lambda_3} \frac{M_{Pl}^2}{\xi} \right)^2
+ \frac{1}{4} \left( \lambda_1 - \frac{\lambda^2_2}{\lambda_3} \right) \frac{M_{Pl}^4}{\xi^2},
\label{Higgs}
\end{eqnarray}
which determines a vacuum expectation value (VEV):
\begin{eqnarray}
\langle \Phi^\prime \rangle = \sqrt{\frac{|\lambda_2|}{\lambda_3} \frac{M_{Pl}^2}{\xi}}.
\label{VEV}
\end{eqnarray}
Expanding as $\Phi^\prime = \langle \Phi^\prime \rangle + \tilde \Phi^\prime$ with $\tilde \Phi^\prime$
being a quantum fluctuation, we have 
\begin{eqnarray}
{\cal{L}} &=& \sqrt{-g^\prime} \Biggl[ \frac{M_{Pl}^2}{2} R^\prime - \frac{1}{2} g^{\prime \mu\nu} 
\partial_\mu \sigma \partial_\nu \sigma - \frac{1}{2} g^{\prime \mu\nu} \partial_\mu \tilde \Phi^\prime 
\partial_\nu \tilde \Phi^\prime - \frac{1}{2} m^2_\Phi \tilde \Phi^{\prime 2} 
\nonumber\\
&-& \frac{\zeta}{M_{Pl}} g^{\prime \mu\nu} \tilde \Phi^\prime \partial_\mu \tilde \Phi^\prime 
\partial_\nu \sigma
- \frac{\zeta^2}{2 M_{Pl}^2} g^{\prime \mu\nu} \tilde \Phi^{\prime 2} \partial_\mu \sigma \partial_\nu \sigma
- \sqrt{\frac{\lambda_3}{2}} m_\Phi \tilde \Phi^{\prime 3} - \frac{\lambda_3}{4} \tilde \Phi^{\prime 4} 
- \frac{\lambda_1}{4} \frac{M_{Pl}^4}{\xi^2} \Biggr],
\label{E-model2}
\end{eqnarray}
where we have simplified the equations by using the relation $| \lambda_2 | \ll \lambda_1, \lambda_3 \approx
{\cal {O}} (0.1)$ and we have defined $m_\Phi = \sqrt{\frac{2 |\lambda_2|}{\xi}} M_{Pl}$. 

As is obvious from (\ref{E-model2}), the SSB of scale symmetry has occurred and as a result the scalar field 
$\tilde \Phi^\prime$ becomes massive while the ``dilaton'' $\sigma$ remains massless, which is nothing but 
a Nambu-Goldstone field. Also notice that the dilaton couples to the scalar field $\tilde \Phi^\prime$ with
derivatives which is one of characteristic features of the dilaton. To establish that $\sigma$ really plays a role
of the Nambu-Goldstone field, it is useful to derive the $\textit{dilatation current}$ associated with scale invariance, 
for which the scale factor $\Omega$ becomes a constant independent of the coordinates $x^\mu$. It is then 
convenient to consider an infinitesimal transformation given by
\begin{eqnarray}
\Omega = e^\Lambda,
\label{Infi-scale}
\end{eqnarray}
where $| \Lambda | \ll 1$. Using the Lagrangian density (\ref{J-model}) and the infinitesimal scale transformation
(\ref{Conformal transf}) with (\ref{Infi-scale}), we find that via the Noether theorem the dilatation
current $J^\mu$ reads
\begin{eqnarray}
J^\mu = \frac{1}{\zeta^2} \sqrt{-g} g^{\mu\nu} \partial_\nu \left( \varphi + \frac{\zeta^2}{2} \Phi^2 \right).
\label{dilatation}
\end{eqnarray}
The dilatation current is certainly conserved 
\begin{eqnarray}
\partial_\mu J^\mu = \frac{1}{\zeta^2} \sqrt{-g} \Box \left( \varphi + \frac{\zeta^2}{2} \Phi^2 \right)
= 0,
\label{Cons-dilatation}
\end{eqnarray}
where we have used the equation (\ref{Derived-eq}). In the E-frame, this current can be written as
\begin{eqnarray}
J^\mu = \frac{1}{2} \sqrt{-g^\prime} g^{\prime \mu\nu} \left[ \frac{2 M_{Pl}}{\zeta} \partial_\nu \sigma
+ \left( \partial_\nu + \frac{2 \zeta}{M_{Pl}} \partial_\nu \sigma \right) \Phi^{\prime 2} \right].
\label{E-dilatation}
\end{eqnarray}
Provided that one defines the dilatation charge as $Q = \int d^3 x J^0$, owing to the linear term in 
$\sigma$ its charge fails to annihilate the vacuum $| 0 \rangle$
\begin{eqnarray}
Q | 0 \rangle  \neq 0,
\label{NG}
\end{eqnarray}
which shows that the dilaton $\sigma$ is the Nambu-Goldstone boson arising from the SSB of
scale invariance.

To close this section, let us summarize the scenario of the SSB explained above and comment on its problems. 
We have started with a scale invariant gravitational theory involving two kinds of scalar fields and only
dimensionless coupling constants. In the process of moving from the J-frame to the E-frame,
we had to introduce a dimensional constant, which is the Planck mass in the present context,
to compensate for the mass dimension of the scalar field. This introduction of the Planck mass
has triggered the SSB of scale symmetry. Let us note that in the conventional scenario of the SSB, 
there is a potential inducing the SSB whereas we have no such a potential in the present SSB.
Nevertheless, the very presence of a solution with dimensional constants justifies the claim 
that the present scenario of the SSB is also nothing but spontaneous symmetry breakdown.  
Actually, this fact was explicitly verified by the dilatation charge, which does not annihilate the
vacuum due to the presence of a linear dilaton.
  
There are, however, at least two problems in this scenario of the SSB. First, it is impossible 
to apply this scenario for the conformally invariant scalar-tensor gravity, for which we must take
either $\xi = \frac{1}{6}$ and $\epsilon = - 1$ or $\xi = - \frac{1}{6}$ and $\epsilon = 1$, due to
$\zeta^{-2} \equiv 6 + \frac{\epsilon}{\xi} = 0$.  
The second problem arises from the lack of the suitable potential in the sense that
we cannot single out a solution realizing the SSB on the stability argument \cite{Fujii}.
Incidentally, though it might be possible that the cosmological argument would pick up an appropriate VEV 
of a scalar field, it is not plausible that the macroscopic physics like cosmology could determine 
a microscopic configuration such as the VEV.
In order to overcome these problems, by following the Coleman-Weinberg mechanism \cite{Coleman}
we have derived an effective potential showing the SSB from radiative 
corrections of gravitational fields associated with higher derivative terms \cite{Oda3, Oda4, Oda5}. 
However, since we have considered the higher derivative gravity in these articles, we have automatically met 
a serious problem of violating the unitarity. The main purpose in this article is to look for an alternative
mechanism for the SSB of conformal symmetry without violating the unitarity.

\section{Spontaneous symmetry breakdown of local conformal symmetry} 

As mentioned in the last paragraph of the previous section, the conformally invariant scalar-tensor
gravity corresponds to either the case of $\xi = \frac{1}{6}$ and $\epsilon = - 1$ or the one of $\xi = - \frac{1}{6}$ 
and $\epsilon = 1$. The former case implies the positive Newton constant (attractive gravitational force) and 
a ghost-like scalar field whereas the latter one does the negative Newton constant (repulsive force) and 
a normal scalar field. Since we wish to work with a theory which never violates the unitarity, we shall choose
the latter case, i.e., 
\begin{eqnarray}
\xi = - \frac{1}{6}, \qquad \epsilon = 1.
\label{xi}
\end{eqnarray}
For generality of the presentation, we will keep $\xi$ and $\epsilon$ for a while, but these constants are 
in fact specified by Eq.  (\ref{xi}).

Although the argument done in case of scale symmetry cannot be directly applied for the case of conformal 
symmetry, let us first follow it to get some lessons. As in the scale invariant theory, suppose that we make a
conformal transformation (\ref{Conformal transf}) for the Lagrangian density (\ref{J-model}). Under this
transformation, the Lagrangian density (\ref{J-model}) with Eq.  (\ref{xi}) is invariant so its form remains
the same. Next, in order to move to the E-frame, let us take a condition $\phi^\prime = \frac{M_{Pl}}{\sqrt{|\xi|}}$,
which is now regarded as a gauge condition, the so-called Einstein gauge or unitary gauge for conformal symmetry, 
in the conformally invariant scalar-tensor gravity. In this context, it is useful to note that 
in quantum field theory, one must impose a gauge condition from the beginning. This gauge choice, however, 
results in the Einstein-Hilbert action with the negative Newton constant (repulsive force) which we cannot 
accept phenomenologically. 

In order to circumvent this problem, we have the following idea: If we set up a different gauge condition including 
the scalar field for conformal symmetry and this gauge condition produces a potential with a mass scale, 
which together with the conformally invariant potential terms in (\ref{J-model}), a new VEV in the total potential 
might yield the positive Newton constant. In this section, we will pursue this possibility and see that this is indeed 
the case. Some people might suspect that a gauge fixing condition for conformal invariance would produce 
such an important physical result, i.e., a change of the sign of the Newton constant. But this suspection is not correct.
The point is that our idea amounts to finding a new VEV in a total potential, which is composed of the potential
in a classical action and the potential coming from the gauge fixing term. This total potential term is obviously 
not BRST-trivial, so our procedure of finding a new VEV in the total potential has a chance of yielding a physically 
meaningful result.    
 
As for a more suitable gauge fixing condition for conformal symmetry, following our previous
study \cite{Oda2}, let us adopt a more general gauge condition
\begin{eqnarray}
a R + b \phi^2 = k,
\label{General gauge}
\end{eqnarray}
where $a, b$ are dimensionless constants while $k$ is a constant having dimensions of mass squared.
Before implementing the gauge fixing procedure, we are ready to present the BRST transformation for conformal
symmetry given by
\begin{eqnarray}
\delta_B g_{\mu\nu} &=& 2 c g_{\mu\nu}, \qquad
\delta_B \sqrt{- g} = 4 c \sqrt{- g}, \qquad
\delta_B R = - 2 c R - 6 \Box c, 
\nonumber\\
\delta_B \phi &=& - c \phi, \qquad
\delta_B \Phi = - c \Phi, \qquad
\delta_B \bar c = i B, \qquad
\delta_B B = \delta_B c = 0,
\label{BRST1}
\end{eqnarray}
where $c, \bar c$ and $B$ represent the FP ghost, the FP antighost and the Nakanishi-Lautrup field, respectively. 
It turns out that all the BRST transformations are nilpotent.

Then, the Lagrangian density for the gauge fixing term and the FP ghost term is given by \cite{Kugo}
\begin{eqnarray}
{\cal{L}}_{GF + FP} &=& - i \delta_B \left[ \sqrt{- g} \bar c \left( a R + b \phi^2 - k + \frac{1}{2} \alpha B \right) \right]
\nonumber\\
&=& \sqrt{- g} \left[  B_* ( a R + b \phi^2 - k ) + \frac{1}{2} \alpha B_* ^2 - 6 i a \bar c \left( \Box 
+ \frac{k}{3 a} \right) c   \right],
\label{GF+FP1}
\end{eqnarray}
where $\alpha$ is a gauge parameter and we have defined $B_* \equiv B + 2 i \bar c c$.
After performing the integration over $B_*$, we have
\begin{eqnarray}
{\cal{L}}_{GF + FP} = \sqrt{- g} \left[  - \frac{1}{2 \alpha} ( a R + b \phi^2 - k )^2 - 6 i a \bar c 
\left( \Box + \frac{k}{3 a} \right) c  \right].
\label{GF+FP2}
\end{eqnarray}
Adding (\ref{GF+FP2}) to the classical Lagrangian density (\ref{J-model}), we obtain a quantum Lagrangian
density given by
\begin{eqnarray}
{\cal{L}}_q &=& \sqrt{-g} \Biggl[ \Biggl( \frac{\xi}{2} - \frac{ab}{\alpha} \Biggr) \phi^2 R 
+ \frac{a k}{\alpha} R - \frac{a^2}{2 \alpha} R^2 - \frac{1}{2} \epsilon g^{\mu\nu} \partial_\mu \phi
\partial_\nu \phi - \frac{1}{2} g^{\mu\nu} \partial_\mu \Phi \partial_\nu \Phi 
\nonumber\\
&-& \frac{\lambda_1}{4} \phi^4 - \frac{\lambda_2}{2} \phi^2 \Phi^2 - \frac{\lambda_3}{4} \Phi^4
- \frac{1}{2 \alpha} ( b \phi^2 - k )^2 - 6 i a \bar c \Biggl( \Box + \frac{k}{3 a} \Biggr) c  \Biggr],
\label{Q-Lag}
\end{eqnarray}
Let us notice that the gauge fixing condition has given rise to not only a potential term but also the terms
involving the curvature scalar. This fact is important in obtaining a physically plausible theory as seen
shortly.
   
Here, to avoid a ghost, the coefficient in front of an $R^2$ term must be positive \cite{Sotiriou}, so we will
take the gauge parameter $\alpha$ to be
\begin{eqnarray}
\alpha = - \frac{a^2}{2}.
\label{Gauge parameter}
\end{eqnarray}
Next, let us focus our attention to a potential to find a VEV
\begin{eqnarray}
V(\phi, \Phi) &=& \frac{\lambda_1}{4} \phi^4 + \frac{\lambda_2}{2} \phi^2 \Phi^2 + \frac{\lambda_3}{4} \Phi^4
- \frac{1}{a^2} ( b \phi^2 - k )^2
\nonumber\\
&=& \frac{\lambda_3}{4} \left( \Phi^2 - \frac{|\lambda_2|}{\lambda_3} \phi^2 \right)^2
+ \frac{1}{4} \left( \lambda_1 - \frac{4 b^2}{a^2} \right)
\left( \phi^2 + \frac{4 b k}{a^2 \lambda_1 - 4 b^2} \right)^2
\nonumber\\
&-& \frac{k^2 \lambda_1}{a^2 \lambda_1 - 4 b^2},
\label{Potential}
\end{eqnarray}
where we used $\lambda_1 - \frac{\lambda_2^2}{\lambda_3} \approx \lambda_1$.
Then, as far as the conditions
\begin{eqnarray}
\lambda_1 > \left( \frac{2b}{a} \right)^2, \qquad
bk <0,
\label{Conditions1}
\end{eqnarray}
are satisfied, for the potential $V(\phi, \Phi)$ there is an absolute minimum at a VEV
\begin{eqnarray}
\langle \Phi \rangle = \sqrt{\frac{| \lambda_2 |}{\lambda_3}} \langle \phi \rangle,
\qquad
\langle \phi \rangle = \sqrt{\frac{4 b k}{- a^2 \lambda_1 + 4 b^2}}.
\label{VEV}
\end{eqnarray}

In order to find the conditions needed for getting a consistent theory without ghosts (except for
the FP ghosts), let us write down the Lagrangian density in the lowest level of approximation 
in the sense that we neglect quantum fluctuations around the minimum (\ref{VEV}). 
Substituting (\ref{VEV}) into (\ref{Q-Lag}) leads to a form
\begin{eqnarray}
{\cal{L}}^{(0)}_q = \sqrt{-g} \Biggl[ \frac{2 k ( a \lambda_1 + \xi b )}{- a^2 \lambda_1 + 4 b^2} R
+ R^2 +  \frac{k^2 \lambda_1}{a^2 \lambda_1 - 4 b^2}
- 6 i a \bar c \Biggl( \Box + \frac{k}{3 a} \Biggr) c \Biggr],
\label{B-Q-Lag}
\end{eqnarray}
where the ghost sector is left for convenience though it is in essence in quantum fluctuations.
First, it is natural to make the coefficient in front of an $R$ term coincide with the square of
the Planck mass divided by the factor $2$:
\begin{eqnarray}
\frac{2 k ( a \lambda_1 + \xi b )}{- a^2 \lambda_1 + 4 b^2} = \frac{M^2_{Pl}}{2} > 0,
\label{Planck}
\end{eqnarray}
which can be achieved by selecting the constant $k$ like
\begin{eqnarray}
k = - \frac{a^2 \lambda_1 - 4 b^2}{4 ( a \lambda_1 + \xi b )} M^2_{Pl}.
\label{k}
\end{eqnarray}

Next, let us look at the ghost sector whose normal form is given by
${\cal{L}} \sim - i \bar c ( \Box - m^2_{FP} ) c$.  Comparing the normal form with
the ghost term in Eq. (\ref{B-Q-Lag}), one must require the following conditions to be satisfied
\begin{eqnarray}
a > 0, \qquad  m^2_{FP} = - \frac{k}{3a} > 0.
\label{Ghost}
\end{eqnarray}
Eventually, using Eqs. (\ref{Conditions1}), (\ref{Planck}) and  (\ref{Ghost}), the parameters
in our theory must satisfy the conditions:
\begin{eqnarray}
a > 0, \qquad  b > 0, \qquad k < 0, \qquad \lambda_1 > \frac{b}{6a},
\qquad  \lambda_1 > \left( \frac{2 b}{a} \right)^2,
\label{F-conditions}
\end{eqnarray}
where we put $\xi = - \frac{1}{6}$.  As a result, starting with the classical Lagrangian density 
(\ref{J-model}) with Eq. (\ref{xi}), which is invariant under conformal transformation 
(\ref{Conformal transf}), performing the BRST quantization by the gauge condition (\ref{General gauge}),
we have reached the Starobinsky model \cite{Starobinsky} with a negative cosmological constant as 
a quantum Lagrangian density in the lowest approximation
\begin{eqnarray}
{\cal{L}}^{(0)}_q = \sqrt{-g} \left[ \frac{M^2_{Pl}}{2} ( R - 2 \Lambda ) + R^2 \right],
\label{Starobinsky}
\end{eqnarray}
where we have ignored the FP ghosts and the negative cosmological constant $\Lambda$ is given by
\begin{eqnarray}
\Lambda = - \frac{\lambda_1 ( a^2 \lambda_1 - 4 b^2)}{16 ( a \lambda_1 - \frac{1}{6} b )^2} M^2_{Pl}.
\label{CC}
\end{eqnarray}

Finally, we would like to move on to quantum fluctuations around the VEV in Eq. (\ref{VEV}) 
by expanding as
\begin{eqnarray}
\Phi = \langle \Phi \rangle + \tilde \Phi,  \qquad \phi = \langle \phi \rangle + \tilde \phi,
\label{Fluctuations}
\end{eqnarray}
where $\tilde \Phi$ and $\tilde \phi$ denote the quantum fluctuations. After a straightforward
calculation, we find that using Eq. (\ref{xi}) the quantum Lagrangian density reads
\begin{eqnarray}
{\cal{L}}_q &=& \sqrt{-g} \Biggl[ \frac{M^2_{Pl}}{2} ( R - 2 \Lambda ) + R^2
+ \left( - \frac{1}{12} + \frac{2 b}{a} \right) ( 2 \langle \phi \rangle \tilde \phi + \tilde \phi^2 ) R
\nonumber\\
&-& \frac{1}{2} g^{\mu\nu} \partial_\mu \tilde \phi \partial_\nu \tilde \phi 
- \frac{1}{2} m^2_\phi \tilde \phi^2 
- \frac{1}{2} g^{\mu\nu} \partial_\mu \tilde \Phi \partial_\nu \tilde \Phi
- \frac{1}{2} m^2_\Phi \tilde \Phi^2
- \frac{1}{2} \lambda_2 \tilde \phi^2 \tilde \Phi^2 
- \frac{\lambda_3}{4} \tilde \Phi^4
\nonumber\\
&-& \left( \lambda_1 - \frac{4 b^2}{a^2} \right) \left( \langle \phi \rangle \tilde \phi^3
+ \frac{1}{4} \tilde \phi^4 \right)  -  i \hat{\bar c} \Biggl( \Box - m^2_{FP} \Biggr)  
 \hat c \Biggr],
\label{Exp-Q-Lag}
\end{eqnarray}
where we have neglected the terms including $\frac{|\lambda_2|}{\lambda_3} \ll 1$ and defined
\begin{eqnarray}
m^2_\phi &=& 2 \left( \lambda_1 - \frac{4 b^2}{a^2} \right) \langle \phi \rangle^2
= \frac{2 b \left( \lambda_1 - \frac{4 b^2}{a^2} \right) }{a \lambda_1 - \frac{1}{6} b} M^2_{Pl}, 
\nonumber\\
m^2_\Phi &=& 2 |\lambda_2| \langle \phi \rangle^2
= \frac{2 b |\lambda_2|}{a \lambda_1 - \frac{1}{6} b} M^2_{Pl}, 
\nonumber\\
\hat{\bar c} &=& \sqrt{6 a} \bar c, \qquad
\hat c = \sqrt{6 a} c, \qquad m^2_{FP} = - \frac{k}{3a} 
= \frac{a^2 \lambda_1 - 4 b^2}{12 a( a \lambda_1 - \frac{1}{6} b )} M^2_{Pl}.
\label{Def-m}
\end{eqnarray}

Here we should comment on the physical meaning of the quantum Lagrangian density Eq. (\ref{Exp-Q-Lag}).
Let us first consider the Nambu-Goldstone theorem. In the theory under consideration, conformal symmetry 
is spontaneously broken so a massless Nambu-Goldstone (NG) boson should emerge in the mass spectrum, 
but at first sight there seems to be no such a massless boson in the Lagrangian density Eq. (\ref{Exp-Q-Lag}). 
With respect to this issue, let us recall that using a scalar field $\omega (x)$ an $R^2$ term can be rewritten as
\begin{eqnarray}
\int d^4 x \sqrt{-g} R^2 = \int d^4 x \sqrt{-g} ( \omega R - \frac{1}{4} \omega^2 ).
\label{R^2}
\end{eqnarray}
It would be this massless scalar field $\omega$ that is a massless NG boson. To put differently, the massless NG boson,
which appears as result of the SSB of conformal symmetry according to the NG theorem, would be absorbed into 
the curvature scalar, thereby an $R^2$ term being induced in Lagrangian density Eq. (\ref{Exp-Q-Lag}).
This new degree of freedom is implicitly supplied by the gauge condition (\ref{General gauge}) since
this gauge condition is second-order in derivatives ($R$ includes the term such as $\sim \partial^2 g$).
Thus, before and after the SSB, the number of dynamical degrees of freedom is unchanged.
It is worth mentioning that a similar but inverse phenomenon has occurred in a Weyl's quadratic gravity
\cite{Oda2}: Starting with an $\tilde R^2$ term in the Weyl geometry, and setting up a gauge condition 
for the Weyl gauge symmetry, we have been able to obtain the Einstein-Hilbert term and a massive 
Weyl gauge field, which ``eats'' a massless NG boson corresponding to the above scalar field $\omega(x)$.

However, at present it seems to be difficult to show the above-mentioned phenomenon explicitly. The reason is that
when we analyze the mass spectrum in the SSB, we usually make use of the unitary gauge where $\phi(x) = const.$
but in the theory at hand, the gauge condition (\ref{General gauge}) plays a critical role so that taking the unitary 
gauge does not make sense. One method for clarifying the present issue is to use a residual symmetry.
In fact, the gauge condition (\ref{General gauge}) has a residual symmetry as long as the infinitesimal parameter
$\Lambda(x)$ for conformal transformation is satisfied with an equation
\begin{eqnarray}
\Box \Lambda + \frac{k}{3a} \Lambda = 0.
\label{Res-symmetry}
\end{eqnarray}
Then, with the infinitesimal global parameters $\alpha_\mu$ and $\beta$ this equation is solved to be
\begin{eqnarray}
\Lambda = \beta e^{\alpha_\mu x^\mu},
\label{Res-symmetry2}
\end{eqnarray}
where $\alpha_\mu$ must satisfy a relation $\alpha_\mu \alpha^\mu + \frac{k}{3a} = 0$. 

By following Ref. \cite{Terao}, this residual symmetry can be used to show that a massless NG boson 
really exists in $g_{\mu\nu}$. According to the Noether theorem, one can construct the Noether
currents $j_\mu \, ^\nu, j^\nu$ and their Noether charges 
$Q_\mu = \int d^3 x j_\mu \, ^0, Q = \int d^3 x j^0$ for global vector $\alpha_\mu$ and scalar $\beta$
parameters, respectively. Then, the residual gauge transformation with the parameters in 
Eq. (\ref{Res-symmetry2}) implies that
\begin{eqnarray}
\delta g_{\mu\nu} = [ i ( \alpha_\rho Q^\rho + \beta Q ), g_{\mu\nu} ] = 2 \Lambda g_{\mu\nu}
\approx 2 \beta g_{\mu\nu},
\label{Res-gauge transf}
\end{eqnarray}
from which we obtain that 
\begin{eqnarray}
[ i Q^\rho, g_{\mu\nu} ] = 0, \qquad
[ i Q, g_{\mu\nu} ] = 2 g_{\mu\nu}.
\label{Res-gauge transf2}
\end{eqnarray}
Hence, we can arrive at a result
\begin{eqnarray}
\langle 0| [ i Q, g_{\mu\nu} ] |0 \rangle \ne 0,
\label{Res-gauge transf3}
\end{eqnarray}
since $\langle 0| g_{\mu\nu} |0 \rangle = \eta_{\mu\nu}$ with $\eta_{\mu\nu}$ being a flat Minkowski
metric. This equation means that the symmetry corresponding to the scalar charge $Q$ is necessarily 
broken, and consenquently $g_{\mu\nu}$ must contain a massless NG mode associated with
the SSB of conformal symmetry.
 
Another peculiar feature of the Lagrangian density Eq. (\ref{Exp-Q-Lag}) is that since the gauge condition 
(\ref{General gauge}) is second-order in derivatives, the FP ghost acquires a kinetic term and has a mass 
which is of order the Planck mass. This situation should be contrasted with that of a Weyl's quadratic
gravity \cite{Oda2} where the FP ghost is non-dynamical. This emergence of dynamics of the FP ghost 
is related to the appearance of a NG boson. Moreover, as seen in Eq.  (\ref{Def-m}), the scalar field $\phi$ 
acquires a mass of the Planck scale while the other scalar field $\Phi$ does a smaller mass than that 
of $\phi$ owing to the existence of the tiny coupling constant $\lambda_2$. 

As a final comment, we should stress that the quantum Lagrangian density Eq. (\ref{Exp-Q-Lag}) is free from 
a ghost problem associated with the scalar field $\phi$ and the Einstein-Hilbert term with a positive 
Newton constant is generated via the SSB of conformal symmetry. Thus, in this sense the $\textit{fake symmetry}$ 
in the conformally invariant scalar-tensor gravity is not really fake when there is an additional scalar field and
the SSB occurs.

\section{Conclusions}

In this article, we have investigated spontaneous symmetry breakdown (SSB) of both global and local
scale invariances. A global scale symmetry has been recently attracted much attention in particle physics
such as BSM \cite{Bardeen} and cosmology such as Higgs inflation \cite{Rubio}. Our findings in this article 
would shed some light on the development of those fields.
 
However, the global scale symmetry is supposed to be broken in the presence of gravitation due to 
no-hair theorems of black holes \cite{MTW}, so it should be promoted to a local scale symmetry, that is, 
conformal symmetry. Then, the important question is how we can break conformal symmetry spontaneously,
which might be related to a resolution to the gauge hierarchy problem and the cosmological constant problem. 
In order to understand the mechanism of the SSB of conformal symmetry, we have examined the conformally
invariant scalar-tensor gravity with two kinds of scalar fields as a simpler model among various conformally invariant
models. Our strategy for the SSB is to first gauge-fix conformal invariance by a suitable condition in a such way 
that a resultant action involving the gauge fixing condition provides us with a nontrivial  potential having 
a mass scale as well as an action of scalar-tensor gravity. The nontrivial potential, together with the conformally
invariant potentials existing in a classical action, determines a new VEV of the total potential, thereby 
producing the Einstein-Hilbert action with a positive Newton constant and an $R^2$ term in the lowest
level of approximation.   
  
As future's works, we wish to apply our conjecture of the origin of the Higgs potential to the other
conformally invariant gravitational theories and construct more realistic models of the BSM. Moreover,
based on these models we wish to consider the gauge hierarchy problem and the cosmological
constant problem etc.


\end{document}